\RequirePackage{fix-cm}
\RequirePackage{amsmath}
\documentclass[twocolumn]{svjour3}          
\smartqed  
\usepackage{graphicx}
\usepackage{graphicx}
\usepackage{algorithm}	
\usepackage{algorithmic}
\usepackage{verbatim}
\usepackage{amsmath}
\usepackage{setspace}
\journalname{Wireless Networks}

\begin{document}

\title{Reducing the Total Cost Of Ownership in Radio Access Networks by Using Renewable Energy Resources
\thanks{ This work is supported by the Turkish State Planning Organization (DPT) under the TAM Project, number 2007K120610.}
}


\author{Turgay Pamuklu \and
        Cem Ersoy 
}


\institute{Turgay Pamuklu \at
              NETLAB, Department of Computer Engineering, Bogazici University, Bebek 34342, Istanbul, Turkey \\
              Tel.: +905366754619\\
              \email{turgay.pamuklu@boun.edu.tr}            \\
	    Cem Ersoy \at
              NETLAB, Department of Computer Engineering, Bogazici University, Bebek 34342, Istanbul, Turkey \\
              Tel.: +902123596861\\
              \email{ersoy@boun.edu.tr}           \\
}

\date{Received: date / Accepted: date}
\maketitle
\begin{abstract}
Increasing electricity prices motivates the mobile network operators to find new energy-efficient solutions for radio access networks (RANs). In this study, we focus on a specific type of RAN where the stand-alone solar panels are used as alternative energy sources to the electrical grid energy. First, we describe this hybrid energy based radio access network (HEBRAN) and formulate an optimization problem which aims to reduce the total cost of ownership (TCO) of this network. Then, we propose a framework that provides a cost-efficient algorithm for choosing the proper size for the solar panels and batteries of a HEBRAN and two novel switch on/off algorithms which regulate the consumption of grid electricity during the operation of the network. In addition, we create a reduced model of the HEBRAN optimization problem to solve it in a Mixed Integer Linear Programming (MILP) Solver. The results show that our algorithms outperform the MILP solution and classical switch on/off methods. Moreover, our findings show that migrating to a HEBRAN system is feasible and has cost-benefits for mobile network operators.
\end{abstract}

\section{Introduction}
Recent reports show that the worldwide electricity production increases each year to satisfy the global energy demands \cite{NathalieDesbrosses2011}. However, the major part of this production is based on fossil fuels which have harmful effects both on economy and environment \cite{Pachauri2007}. The Economy struggles with the increasing of electricity prices \cite{P.CaprosA.DeVitaN.Tasios2013} and depleting of fossil fuels \cite{quaschning2009renewable} and the environment suffers from the greenhouse gases that mostly come from the combustion of these fossil fuels. Therefore economic and environmental problems lead to a renewed interest in related research to reduce the consumption of fossil fuels and on-grid electricity in many different areas. The radio access network (RAN) researchers are not an exception, they are also motivated to explore energy-efficient solutions to reduce the on-grid energy consumption.
\par A considerable amount of these studies focus on providing a more energy-efficient operating method for the base stations when they aim to achieve reducing the energy consumption in a RAN. They have two important motivations: the base stations in a RAN form the main source (60-80\%) of the energy consumption \cite{oh2011toward} and they are under-utilized during most of their operating time. Improvement efforts usually take the advantage of data traffic changes in time \cite{Wu2015}. Some of the researchers adapt the transmission power of the base stations by the traffic variation, which is called cell-breathing \cite{Pamuklu2013,Yigitel2014} and the others use an even more radical version of this method, which selectively switches off the base stations according to the traffic fluctuations \cite{oh2011toward,son2011energy,Yildiz2013,Zhang2013}. The results show that both techniques reduce a significant amount of energy consumption in a RAN. In this paper, we also use the latter technique as a tool in our solution to deal with the environmental and economic effects of the excessive use of on-grid electricity.
\begin{figure}
\centering
\includegraphics[width=0.48\textwidth]{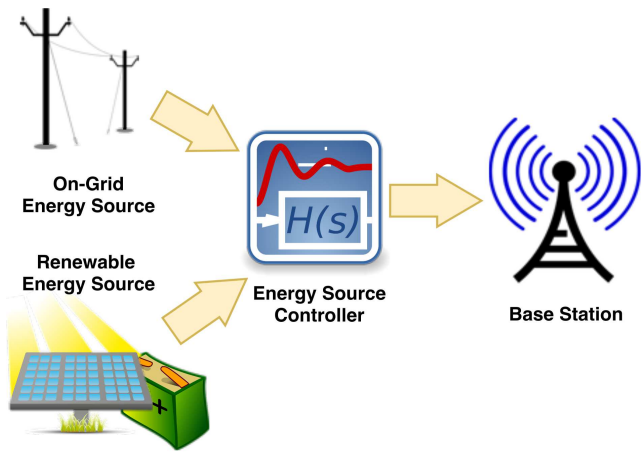}
\caption{\label{fig:renewable_system} Energy System Model in a HEBRAN.}
\end{figure}
\par Another key component we adopt in our solution is the use of renewable energy to reduce the consumption of on-grid electricity. In our solution, a base station has two energy sources complementing each other (Figure~\ref{fig:renewable_system}). The renewable energy source reduces the electrical grid energy consumption and the electrical grid provides the energy demand in the case of the lack of the insufficient renewable energy. This type of network is named as hybrid energy based radio access network (HEBRAN) \cite{Hassan2013}. Although operating this type of a network is a new concept, there has been significant amount of study focusing on the problems of this new kind of network. For example, Han et al. \cite{hanoptimizing} demonstrate that operating a system with hybrid energy sources is an NP-hard problem and they decompose this problem into two sub-problems to handle the complexity. In the first sub-problem, they aim to optimize the allocation of green energy (which is generated by a solar panel) to each time interval for each base station. In the second subproblem, they change the transmission power of the base stations in each time interval by considering their allocated green energy. Therefore they use the generated green energy more efficiently to reduce the overall energy consumption. Carreno et al. \cite{Carreno2013} analyze a single base station system which can change its coverage area with the amount of the renewable energy in this base station. The critical results in this study show a cross-correlation between the user satisfaction and the ratio of the renewable energy used in this base station. Farooq et al. \cite{Farooq2017} propose an energy sharing framework, in which base stations in the network can share their renewable energy over the grid. Sheng et al. also aim to reduce the energy consumption by using the centralized and distributed algorithms that get benefit from smart grid architecture \cite{Sheng2017}.
\par The third key component in our study is that we provide a system-wide solution to reduce the total cost of ownership (TCO) of a HEBRAN. This kind of a network has two main expenditures: the capital expenditure which is the investment cost of constructing this new system and the operational expenditure which is the cost of maintaining this new system \cite{johansson2007cost}. In detail, the capital expenditure of our hybrid system has two main components: cost of the solar panels and cost of the batteries in which the panels and the batteries are installed on each base station in the network. Both the installation cost and the amount of harvested renewable energy of a solar panel linearly increase by its panel size. Also, the cost of a battery and its energy storing capacity correlate by each other. Therefore we have a renewable system sizing problem which can be briefly summarized as the selection of the most cost-efficient size of the solar panels and the batteries in a HEBRAN. On the other hand, the operational expenditure of this type of network has only one main component: the electrical grid cost that is used to operate the base stations of this network. The main tool we select to cut down this second expenditure is the base station switch on/off techniques with considering the amount of stored renewable energy in the base stations. In addition to the electrical grid cost, maintenance cost of the solar panels and the batteries can be added as an operational cost. In summary, in this paper, we propose an optimized solution to reduce the total cost of ownership (TCO) of a hybrid energy based radio access network (HEBRAN).
\par To the best of our knowledge, our paper is the first study which aim to reduce the TCO of a RAN in which all base stations have both renewable and on-grid energy sources. For example, two recent studies \cite{Fan2016,Lee2017} aim to reduce the on-grid energy consumption but they did not consider the capital expenditure of a HEBRAN. Also they focus on a problem which only aim to balance the traffic load between a macro base station and pico\&micro base stations around this macro base station. In a broader study by Han et al. \cite{Han2016}, the authors aim both the sizing problem of a HEBRAN and its energy-efficient operation problem but their RAN also have non-renewable energy base stations and the main idea of their method is offloading the traffic from the hybrid energy base stations to these non-renewable energy base stations to reduce the size of the panels and batteries in the hybrid energy base stations. In their following work \cite{Han2015}, Han et al. propose a similar solution but at that time they offload the traffic from the pure renewable energy base stations to the pure on-grid energy base stations. Our paper has three different aspects from these two papers. First, in our system each base station has their own renewable energy source and has a connection to the electrical grid. Our studies show that deploying a solar panel in each base station provides better results. Second, our operational decisions consider the remaining energy in the batteries of the base stations and focus on using renewable energy as efficient as possible in each base station to reduce the electrical grid energy consumption. On the other hand, Han et al. \cite{Han2016,Han2015} use a predetermined constant parameter for the renewable energy percentage in the network and this approach is not efficient to preserve the renewable energy. Lastly, by using this constant they make a static connection between the operational expenditure and the capital expenditure. On the contrary, our approach dynamically connects these two expenditures to reduce the TCO of the network in an efficient way.
\par The remainder of this paper is organized as follows. We describe this new type of RAN and its cost optimization problem in the second and third section, respectively. In the fourth section, we propose novel solutions to overcome this problem. Fifth section provides the results of these solutions and the last chapter concludes the paper.
\begin{figure}
\centering
\includegraphics[width=0.48\textwidth]{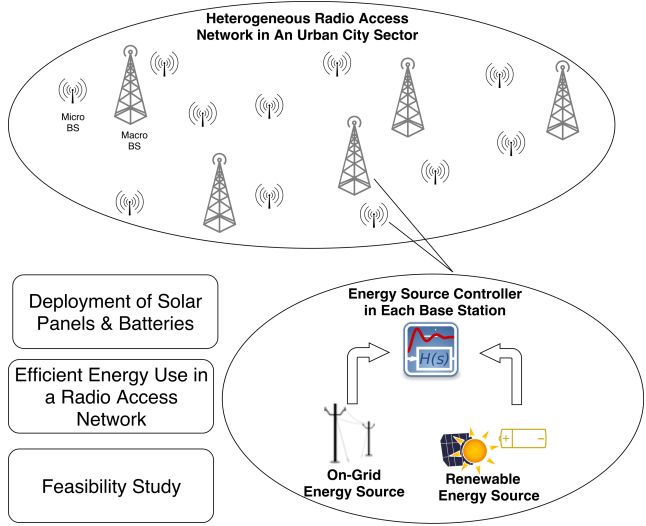}
\caption{\label{fig:overview} The Hybrid Energy Based Radio Access Network Model.}
\end{figure}
\begin{table}
\centering
\caption{\label{tab:Notations} List of Notations}
\begin{tabular}{|c|p{6.5cm}| }
\hline
Sets &  Explanation \\ \hline
$i\in\mathcal{I}$ & set of base stations\\
$j\in\mathcal{J}$ & set of locations\\
$t\in\mathcal{T}$ & set of discrete time intervals\\ \hline
Variables&  Explanation \\ \hline
$s_{i}$ & solar panel size of BS i\newline ($0\leq s \leq 6 | b\in N$ )\\
$b_{i}$ & battery size of BS i \newline ($0\leq b \leq 8 | b\in N$ )\\
$r_{it}$ & renewable energy usage ratio of BS i \newline ($0\leq r_{it} \leq 1 | r_{it}\in \Re$ )\\
$x_{it}$ & base station on/off decision \newline ($x_{it} \in \{0, 1\}$)\\
$z_{ijt}$ & assigning decision \newline ($z_{ijt} \in \{0, 1\}$)\\ \hline
Input &  Explanation \\ \hline
$U_{jt}$ & data traffic demand of location j\\ 
$S_{ij}$ & service rate of BS i to location j  \\
$a^E_{i}$ & energy consumption of BS i (kW)\\
$c^E$ & electrical grid cost (\$/kW)\\
$c^S$ & unit cost of a solar panel (\$/kW) \\
$c^B$ & unit cost of a battery (\$/2.5kW) \\
$a^B$ & unit capacity of a battery (2.5kW)\\
$\rho$ & utilization bound of a base station \\
$G(t)$ & generated renewable energy in a time interval \\
\hline
\end{tabular}
\end{table}    
\section{System Description}
\label{sec:systemDescription}
In this section, we start to describe a HEBRAN system as part of a real-world scenario and then we detail the parts of this system in the following subsections.
\subsection{Scenario}
\label{sec:scenario}
The following scenario is a typical situation that the researchers should aim to solve the cost optimization problem explained in this paper:
\begin{enumerate}
\item A mobile network operator (MNO) has a RAN deployed in an urban city sector. This RAN has both macro and micro type base stations (an heterogeneous network), in which these two base station types  have different maximum transmission power level and energy consumption specifications.
\item The operator wants to supply their base stations with renewable energy sources like solar panels for both improving the energy-efficiency and the sustainability of their RAN.
\item In addition, the base stations in the network are not well utilized, their energy consumption is almost the same for different traffic loads. Therefore, the mobile operator wants a new operational method in order to reduce this idle unnecessary energy consumption. Moreover, they should efficiently combine this technique with the usage of the renewable energy sources.
\item Also they need a feasibility analysis which show the economic efficiency of this new generation RAN in an urban sector.
\end{enumerate}
\par Figure~\ref{fig:overview} demonstrates this scenario. Before starting to formulate the problem in this scenario, we have to detail the critical parts of a HEBRAN in the following subsections. Table~\ref{tab:Notations} summarizes the notations we use in this section. 

\subsection{Traffic Model}
\label{sec:userTraffic}
\begin{figure}
\centering
\includegraphics[width=0.48\textwidth]{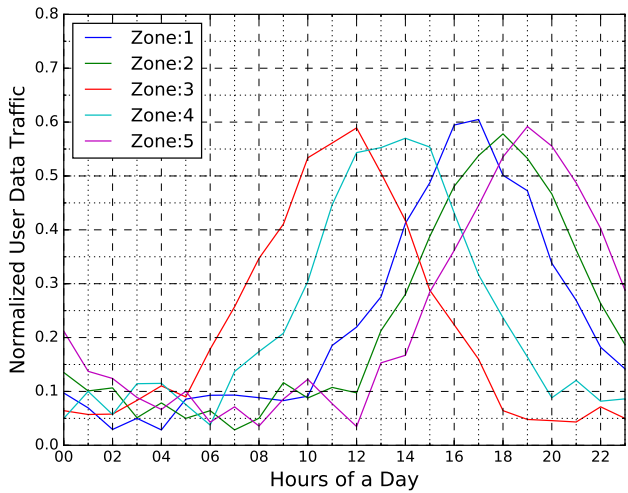}
\caption{\label{fig:traffic_day.pdf} Five different data traffic patterns in a day period.}
\end{figure}   
\par First, we formulate the downlink data traffic. Since this type of traffic rapidly increases by the increasing usage of the internet and the invention of the new wireless network devices. On the other hand, the mobile users need far less bitrate for the uplink traffic \cite{Peng2011}. For that reason, providing an energy-efficient solution for the downlink traffic is more crucial than a solution for the uplink traffic. In addition, the new technologies (beyond 3G) use packet switching technologies for the voice traffic \cite{Poikselka2012}. We should noticed that even we only consider the downlink traffic in this paper, our methods are also applicable for the uplink traffic.   
\begin{figure}
\centering
\includegraphics[width=0.48\textwidth]{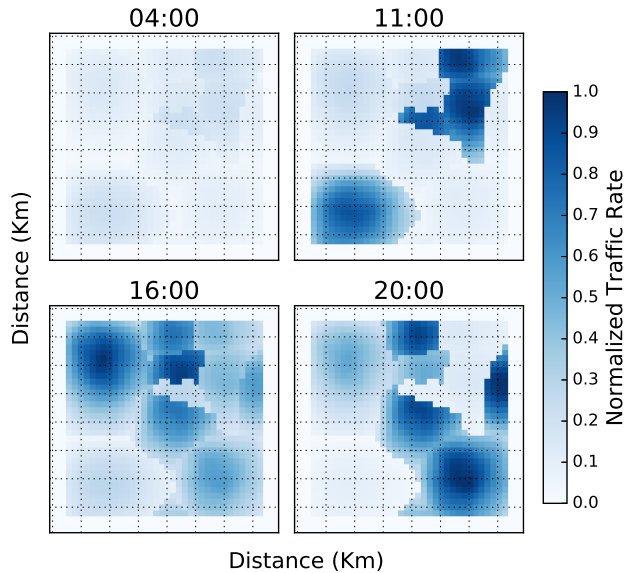}
\caption{\label{fig:traffic_dist_day.png} The traffic rate of our covered sector in a day period.}
\end{figure}
\par One of the most important studies about the dynamics of the data traffic in urban sectors is written by Peng et al. \cite{Peng2011}, in which they analyze two-month long real network data of a real mobile network operator. Their analysis shows that the data traffic significantly changes both temporally and spatially in four different urban regions. In addition, they present two critical findings that inspire our solution in this paper. The first one is that the temporal change of data traffic is more powerful in a day period but it is not significant between the consecutive days. The second is that the data traffic loads are diverse between very close locations especially in their peak hours. For that reasons, although we do not use real data traffic in this paper, we model it by considering these authors findings and make it diverse both temporally and spatially which are explained in the following paragraphs.  
\par The temporal traffic profile we use in this paper evolves from several papers. First, Marsan et al. \cite{Marsan2010} propose a formula which create a one day period sinusoidal shape traffic for a RAN. Then, Hossain et al. \cite{Hossain2010} modify this formula by adding a random fluctuation in a day period. Finally Zhang et al. \cite{Zhang2013} add a multiplier into this formula to create a diversity between different locations. We modify this last formula by adding a fluctuation between the days of a year and try to make it diverse between the weekdays and weekends by considering the findings of Peng et al. \cite{Peng2011}. Our formula is given in Equation~\ref{eq:trafficCreator} in which $\kappa$ is the value of the peak hour traffic which depends on whether it is a weekday or weekend, $\varphi$ is a random value between the $3\pi/4$ and $7\pi/4$ which determines the peak hour of the traffic profile,  $\nu$ determines the abruptness of the traffic profile and $n(t)$ is a random value which provides a fluctuation in this traffic profile. Therefore, we can model the variation of data traffic between each hour and each day by this formula. 
\begin{equation}
\label{eq:trafficCreator}
f_{z}(t) = \frac{\kappa}{2^{\nu}}[1+\sin(\pi t/12 + \varphi )]^{\nu} + n(t)
\end{equation}
\begin{equation}
\label{eq:userTraffic}
U_{j}(t) = f_{H_{j}^{z}(.)}(t),     H_{j}^{z}(.)=\{z|z < 5, z\in N)\}
\end{equation}
\par
\begin{figure}
\centering
\includegraphics[width=0.48\textwidth]{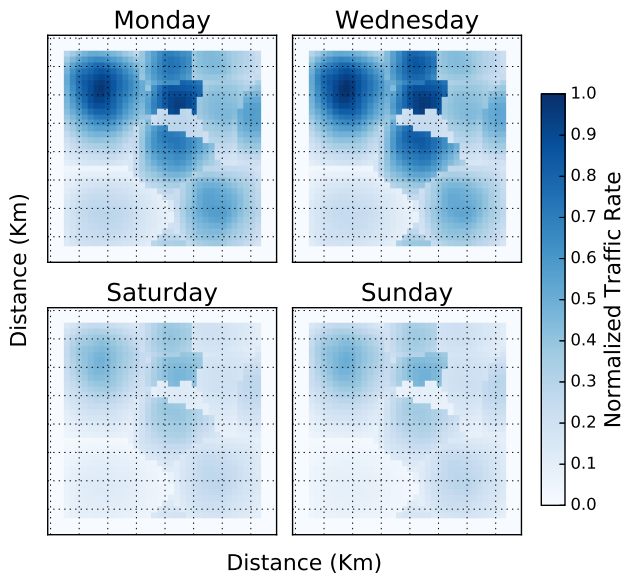}
\caption{\label{fig:traffic_dist_week.png} The traffic rate of our covered sector in a week period.}
\end{figure}

 We use a three-step method to create a traffic diversity between the locations of the urban sector according to the findings in \cite{Peng2011}. In the first step, we create five different traffic profiles by Equation~\ref{eq:trafficCreator} in which each traffic profile have different peak hours (Figure~\ref{fig:traffic_day.pdf}). In the second step, we create a map which have different hotspots by using the kernel density estimation method \cite{Silverman1986}. These hotspots present the different districts of an urban sector such as business or residential districts. In the third step, we assign the traffic profiles which we create in the first step to these hotspots by using a pre-calculated function $H_{j}^{z}(.)$ which maps the $j$ location to the $zth$ traffic profile which is shown in Equation~\ref{eq:userTraffic}. Therefore we create a spatial diversity between the five different districts which have different peak hours. Figure~\ref{fig:traffic_dist_day.png} and Figure~\ref{fig:traffic_dist_week.png} show the change of traffic rates in a day period and in a week period for these different districts, respectively. Lastly, we minimize the traffic rates in the edge of the covered sector (white zones) to eliminate the border effect. 

\subsection{Channel Model}
\label{sec:serviceRate}
We consider an urban sector where a set of base stations ($I$) are already deployed to serve a set of location ($J$) who demand a service from these base stations. This section explains how these base stations satisfy the data traffic required by these locations ($U_{jt}$, Section~\ref{sec:userTraffic}). First, we have to calculate the path loss ($L_{ij}$) between each base station ($i \in I$) and each location ($j \in J$) in this sector. For that purpose, we use the macro and micro NLOS path loss models from the ITU-R report \cite{ReportITU-RM.2135-12009} according to the type of the base station ($i \in I$). After calculating the path loss, we calculate the received signal - noise ratio (SNR, $\Gamma_{ij}$) at the location side by Equation~\ref{eq:constSNR}.
\begin{equation}
\label{eq:constSNR}
\Gamma_{ij}=\frac{L_{ij} P^{T}_{i}}{\sigma^{2}}
\end{equation}
\begin{equation}
\label{eq:constSpecEff}
S_{ij}=B\log_{2}(1+\Gamma_{ij} ) (bit/sec)
\end{equation}
In this equation, $L_{ij}$ is the path loss, $P^{T}_{i}$ is the transmission power of the base station $i \in I$  and $\sigma^{2}$ is the Gaussian noise. Dealing with the interferences between the base stations is out of scope in this paper and we assume it is well managed by the frequency planning. Finally, we use the Shannon capacity formula \cite{Shannon1956} to calculate the service rate (spectral efficiency) of a base station for a specific location (Equation~\ref{eq:constSpecEff}). In this equation, $B$ represents the bandwidth allocated to the base station $i \in I$ and $S_{ij}$ represents the service rate of this base station to the location $j \in J$.
\subsection{Base Station Energy Consumption Model}
\par We use two different types of base stations in our study which are named as macro and micro base stations. Auer et al. analyze the energy consumption model of several types of base stations in their paper including these two types of base stations \cite{Auer2011}. According to their findings, both macro and micro base stations have a large static energy consumption even they have not been assigned to any location. With this in mind, we prefer the base station switch on/off method over the cell breathing method in which base stations have different transmission power levels \cite{Wu2015}. Therefore the base stations in our model have two level of energy consumption: the value of their maximum transmission power and the value of their deep sleep state. The deep sleep state energy consumption does not change by the decisions of our techniques and remaining as a constant value in the lifetime of the network so we omit it from the formula but it can be easily added to the TCO as a constant value. We provide the energy consumption values for both macro and micro base stations in Section~\ref{sec:results}.
\subsection{Renewable Energy System Model}
\label{sec:renEnergy}
Figure~\ref{fig:renewable_system} shows an electrical grid energy supported renewable energy system for a base station. In this system, the solar panel harvests the renewable energy from sunlight, the battery stores this energy and the energy source controller chooses the energy source that is used to supply the base station. According to Hassan et al. \cite{Hassan2013a}, itself based on \cite{valerdi2010intelligent}, harvested green energy may be used directly by the system or may be stored in a battery for future use. Storing the green energy in a battery does not only prevent the wasting of the excess green energy but also maintains the more efficient usage of the green energy. Although this system is not common for a base station nowadays, Bloomberg Finance group announced in one of their latest reports that using a battery storage alongside with solar panels will become an ordinary method for a rooftop system in 2020 \cite{Parkinson2014}. Moreover, a recent report about the solar panel technology in Germany also supported this new system. They suggest that the prices of solar panels drop 19\% each year and they mention that the prices continue to drop year by year \cite{Wirth2016}. With this in mind, we directly focus on the base stations which have their own solar panels and batteries. We explain the model of a solar panel and a battery in our system in the following sections. 
\begin{figure}
\centering
\includegraphics[width=0.48\textwidth]{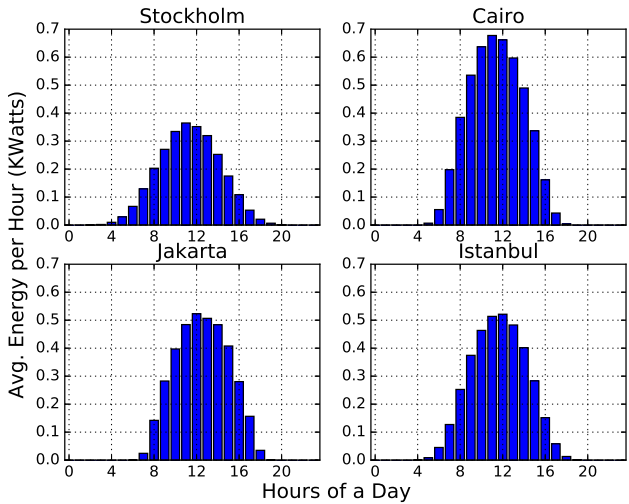}
\caption{\label{fig:harvested_hourly} Distribution of Harvested Solar Radiation in a Day Period.}
\end{figure}  
\subsubsection{Solar Panel Model}
\label{sec:solarEnergy}
\par
\begin{figure}
\centering
\includegraphics[width=0.48\textwidth]{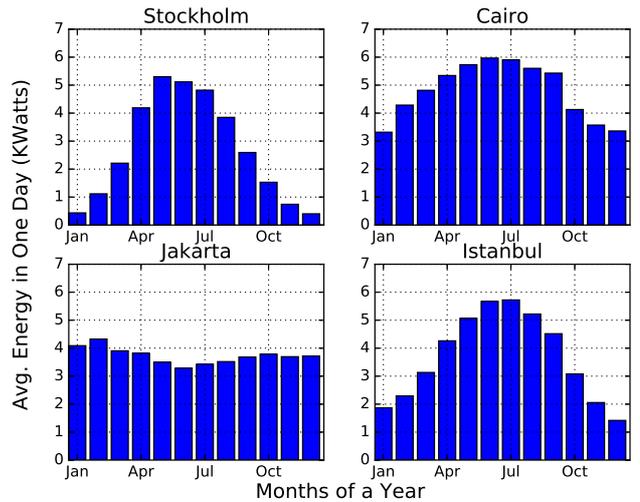}
\caption{\label{fig:harvested_daily} Distribution of Harvested Solar Radiation in a Year Period.}
\end{figure}

 Choosing an appropriate solar panel size\footnote{We have to clarify that in this paper, we use the \textquotedblleft solar panel size\textquotedblright clause to define the energy generating capacity of this panel.} for a base station in a HEBRAN is crucial for both increasing the renewable energy generating capacity and reducing the capital expenditure of this network.The relation between the size of a solar panel and the amount of the generated renewable energy by this panel is linear according to National Renewable Energy Laboratory \cite{NationalRenewableEnergyLaboratory}. While considering this relation, we use the empirical data from the pvWatts application to calculate the amount of the generated energy of a solar panel for each time interval \cite{NationalRenewableEnergyLaboratory}. Their 30 years historical weather data provide us to calculate the detailed solar energy generation rate of a panel ($G(t)$) for different cities. They provide each hour of the day data, thus we can simulate the change of the solar energy in several time scales. Figure~\ref{fig:harvested_hourly} and Figure~\ref{fig:harvested_daily} shows the change of generated energy of a $4kW$ size of solar panel for different cities.
\par The other important thing is that the size of a solar panel directly affects the installation cost (capital expenditure) of a HEBRAN. Equation~\ref{eq:panelCost} shows how to calculate this installation cost, in which $c^S$ is the unit cost of a solar panel $s_{i}$ is the solar panel size of base station $i$. We can use this linear cost model, if the energy harvesting capacity of these panels is lower than 10kWh \cite{NationalRenewableEnergyLaboratory}.
\begin{equation}
\label{eq:panelCost}
PanelCost(\$) = \sum\limits_{\\i\in\mathcal{I}}c^S  (\$/kW) * s_{i} (kW)
\end{equation}
\begin{equation}
\label{eq:solarPanel}
Array Area (m^2) = \frac{Size (kW)}{Efficiency (\%)} * \frac{m^2}{1 kW}
\end{equation}
\par However the solar panel size is not limited by only its installation cost. According to our scenario, we have to install these solar panels near to the base stations which are usually on a rooftop of a building in an urban sector. Therefore we have to limit the size of a solar panel by looking at the array area of this panel. Equation~\ref{eq:solarPanel} shows the relation between the array area and the size of a solar panel \cite{Piro2013,NationalRenewableEnergyLaboratory}. We limit the maximum size of a solar panel as $6kW$ as a result of this relation.

\subsubsection{Battery Model}
\label{sec:batteryModel}
\par First of all, we have to give some explanation about the battery technology we use in this study. We choose the lithium-ion based batteries in light of a recent report published by International Renewable Energy Agency (IRENA) \cite{Grothoff2015}. This report emphasizes that lithium-ion batteries have several advantages when we compare them with the other battery types. Especially their deep discharge cycle capability that helps the base stations to consume high amount of energy in a short time and their power density, which makes them preferable to be used in a small area, encourage us to choose lithium-ion batteries. The same report also advises the lithium-ion cells because of their decreasing price trends. Lastly, the Navigant Research emphasizes that the distributed energy storage system (DESS) market - our system is a kind of DESS - prefers lithium-ion batteries as their first choice for a battery \cite{Eller2016}.
\par One of the purposes of this work is the appropriate battery size selection for an energy-efficient solution. However, the battery size and renewable energy consumption rate has a complicated relation, which increases the complexity of this size selection problem. Han et al. \cite{Han2016} propose the following equation to define this non-linear relation, itself based on \cite{Badawy2010}.
\begin{multline}
\label{eq:renfosawa}
R_{i}(t) = \min\Big\{\max\left\{R_{i}(t-1) + s_{i}G(t)  - a^E_{i} r_{it}, 0\right\}, \\
a^B b_{i} \Big\} \forall i\in\mathcal{I} , \forall t\in\mathcal{T} 
\end{multline}
In this equation, $R_{i}(t)$ is the stored/remaining battery energy of the base station $i \in I$  in time interval $t \in T$. The maximum value of $R_{i}(t)$ is the battery storage capacity value, $a^B b_{i}$, in which $a^B$ is the unit storage capacity and  $b_{i}$ is the size of the battery which is a decision parameter in our objective function. $G(t)$ is the generated renewable energy of a unit size solar panel in the time interval $t$, $s_{i}$ is the size of the solar panel and $a^E_{i}$ is the energy consumption of a base station in a single time interval. Finally, the most important part, $R_{i}(t-1)$ is the remaining battery energy from the previous time interval which means that the $r_{it}$ decisions and $G(t)$ values in the previous time intervals, directly affect the current $r_{it}$ decision. In summary, our problem becomes a non-linear problem due to this relation.
\par In this paper, the price of a lithium-ion battery changes linearly with the capacity of this battery \cite{Lacey2016}. Equation~\ref{eq:batteryCost} shows this relation, in which $c^B$ is the unit battery price and $b_{i}$ is the size of the battery at base station $i$ .The other cost components of a battery system are omitted in this study because they do not change with the size of a battery, but can be added easily as a constant without changing the performance of the proposed solution. In addition, we assume the batteries in our system are charged/discharged linearly in their lifetime for the simplicity. 
\begin{equation}
\label{eq:batteryCost}
BatteryCost(\$) = \sum\limits_{\\i\in\mathcal{I}}c^B  (\$/kW) * b_{i} (kW)
\end{equation}
\section{Cost Optimization Problem of a HEBRAN}
\label{ch:problem}
In the previous section, we describe a scenario in which the renewable energy systems using together with the base stations in an urban sector and we explain the critical parts of this scenario. This section explains the optimization problem which begins by the details of the objective function and the constraints in the problem and finishes with the description of the complexity of this formulated problem. 
\subsection{Problem Formulation}
\label{ch:problemdetails}
\par The optimization problem of a HEBRAN can be given as:
\begin{equation}
\label{eq:obj1}
\min \left\{\sum\limits_{\\i\in\mathcal{I}}c^S  s_{i}+\sum\limits_{\\i\in\mathcal{I}}c^B  b_{i}+\sum\limits_{\\i\in\mathcal{I}}\sum\limits_{\\t\in\mathcal{T}}c^E a^E_{i}(1-r_{it})x_{it}\right\}
\end{equation}
$s.t.$
\begin{equation}
\label{eq:qos}
\sum\limits_{\\i\in\mathcal{I}}S_{ij} z_{ijt} \geq U_{jt} , \forall j\in\mathcal{J}, \forall t\in\mathcal{T} 
\end{equation}
\begin{equation}
\label{eq:capacityLimit}
\sum\limits_{\\j\in\mathcal{J}}\frac{U_{jt}}{S_{ij}}z_{ijt} \leq \rho , \forall i\in\mathcal{I}, \forall t\in\mathcal{T} 
\end{equation}
\begin{equation}
\label{eq:assignMaxOneBS}
\sum\limits_{\\i\in\mathcal{I}}z_{ijt} \leq 1, \forall j\in\mathcal{J} , \forall t\in\mathcal{T} 
\end{equation}
\begin{equation}
\label{eq:shouldBeActive}
x_{it} \left\vert{J}\right\vert - \sum\limits_{\\j\in\mathcal{J}} z_{ijt} \geq 0, \forall i\in\mathcal{I} , \forall t\in\mathcal{T} 
\end{equation}
\begin{equation}
\label{eq:smallra}
r_{it} \leq \min\Big\{ \frac{R_{i}(t)}{a^E_{i}}x_{it}, 1\Big\}, \forall i\in\mathcal{I} , \forall t\in\mathcal{T} 
\end{equation}
\begin{equation}
\label{eq:smallrc}
r_{it} \geq \min\Big\{ \frac{R_{i}(t)}{a^E_{i}}x_{it}, 1\Big\}, \forall i\in\mathcal{I} , \forall t\in\mathcal{T} 
\end{equation}
\par  We aim to minimize the TCO of this system which is the sum of the capital expenditure and the operational expenditure. Migrating to a renewable energy system has two main capital expenditure types which are shown as the first two components in Equation \ref{eq:obj1}. The first one is the installation cost of the solar panels and the second one is the installation cost of the batteries. Since these components are briefly explained in Section~\ref{sec:renEnergy}, we skip the details of the calculation of these components in this section. The third component in Equation \ref{eq:obj1} formulates the on-grid energy consumption, in which $c^E$ is the electrical grid cost per kilowatt in a time interval, $a^E_{i}$ is the energy consumption of a base station in a time interval, $r_{it}$ is the ratio of the renewable energy consumption and $x_{it}$ is the binary decision variable states that the base station $i$ is in switched on (active) mode . This equation shows that we can reduce the operational expenditure by increasing either the renewable energy consumption ratio ($r_{it}$) or decreasing the number of the active base stations ($x_{it}$) in this network. Moreover, we have to notice that the operational expenditure is calculated as the summation of every time interval during the life-cycle of this renewable system. The maintenance cost of this renewable system, which is another component of the operational expenditure, is not included in  Equation~\ref{eq:obj1}. The reason is that this maintenance cost is a constant value and does not change by any decision variable in Equation~\ref{eq:obj1}.
\par We explained how to calculate the traffic rates of the locations in an urban sector in Section~\ref{sec:userTraffic}. To satisfy this data traffic (Quality of Service), each location in this sector should be serviced by a base station at every time interval. Inequality~\ref{eq:qos} formulates this assigning operation, in which $U_{jt}$ is the traffic rate of the location $j$ in the time interval $t$, $S_{ij}$ is the maximum service rate of the base station $i$ to the location $j$ and $z_{ijt}$ is a binary decision variable which equals to one when the base station $i$ is assigned to the location $j$ in the time interval $t$. We have to notice that instead of adding another constraint for the coverage of the region, we choose the minimum $U_{jt}$ larger than zero for each location and the time interval. Since at least one base station should be assigned to a location that have a traffic rate larger than zero, Inequality~\ref{eq:qos} also provides the full coverage of the region. 
\par In addition to satisfying the traffic rates of the locations and providing the coverage of the sector, the delay of service in the base stations should be lower than a reasonable quality of service. Inequality~\ref{eq:capacityLimit} provides this quality of service by limiting the total service load on a base station, in which $\rho$ is the maximum system load that is allowed on a base station in this network.
\par Inequality~\ref{eq:qos} allows that more than one base station may serve to the same user in a location in the same time interval. Although a wireless communication technology that has a Coordinated Multi-Point (CoMP) property may use more than one base station to serve the same user \cite{Irmer2011}, the technology that we study may not support this property. Therefore, we add Inequality~\ref{eq:assignMaxOneBS} to guarantee that a user is served by only one base station in the same time interval. For the CoMP cases, this inequality can be removed.
\par The decision parameter $z_{ijt}$ also determines one of the objective function parameters: the base station switch on/off decision ($x_{it}$). If a base station serves at least one location in a time interval, that base station should not switch off in this time interval. Inequality~\ref{eq:shouldBeActive} defines this relation.
\par As we mentioned before, the base stations in a HEBRAN can be supplied from a renewable energy source and the on-grid energy source at the same time. The ratio of the energy consumption between these two sources is determined by the renewable energy ratio variable ($r_{it}$), which is formulated in Inequality~\ref{eq:smallra} and Inequality~\ref{eq:smallrc}. This value can be maximum one, which means that a base station consumes only renewable energy in this time interval. However, assuming that if the renewable energy of a base station $R_{i}(t)$ is not enough to supply the energy consumption of this base station ($a^E_{i}$) in the duration of a time interval $t$, the renewable energy consumption  and on-grid energy consumption will be equal to the $R_{i}(t)$ and $a^E_{i}-R_{i}(t)$, respectively. In this case the ratio of the renewable energy consumption equals to $\frac{R_{i}(t)}{a^E_{i}}$ .We assume that a base station always prefers to consume the renewable energy over the on-grid energy which is provided by Inequality~\ref{eq:smallrc}. Since the on-grid energy price does not change between the time intervals in our scenario, this assumption does not have a negative effect on minimizing the operational expenditure.
\subsection{The Complexity Analysis}\label{sec:complexityAnalysis}
\par This section briefly explains the NP-complete characteristic of the problem. Let us consider a special case of the problem in which the unit solar panel ($c^S$) and the unit battery ($c^B$) are overpriced that deploying these new equipments does not have any benefit to reduce the TCO. Then the objective function can be simplified as Equation~\ref{eq:obj2}, while the constraints~\ref{eq:qos} - \ref{eq:shouldBeActive} remain the same for this specific scenario.
\begin{equation}
\label{eq:obj2}
\min \left\{\sum\limits_{\\i\in\mathcal{I}}\sum\limits_{\\t\in\mathcal{T}}c^E a^E_{i} x_{it}\right\}
\end{equation}
\par This problem is a best-known base station switching problem for any instance of time interval and it can be reduced to the vertex cover problem \cite{oh2013dynamic,Karp1972}. Therefore the problem in this paper is also an NP-complete problem.
\begin{figure}
\centering
\includegraphics[width=0.48\textwidth]{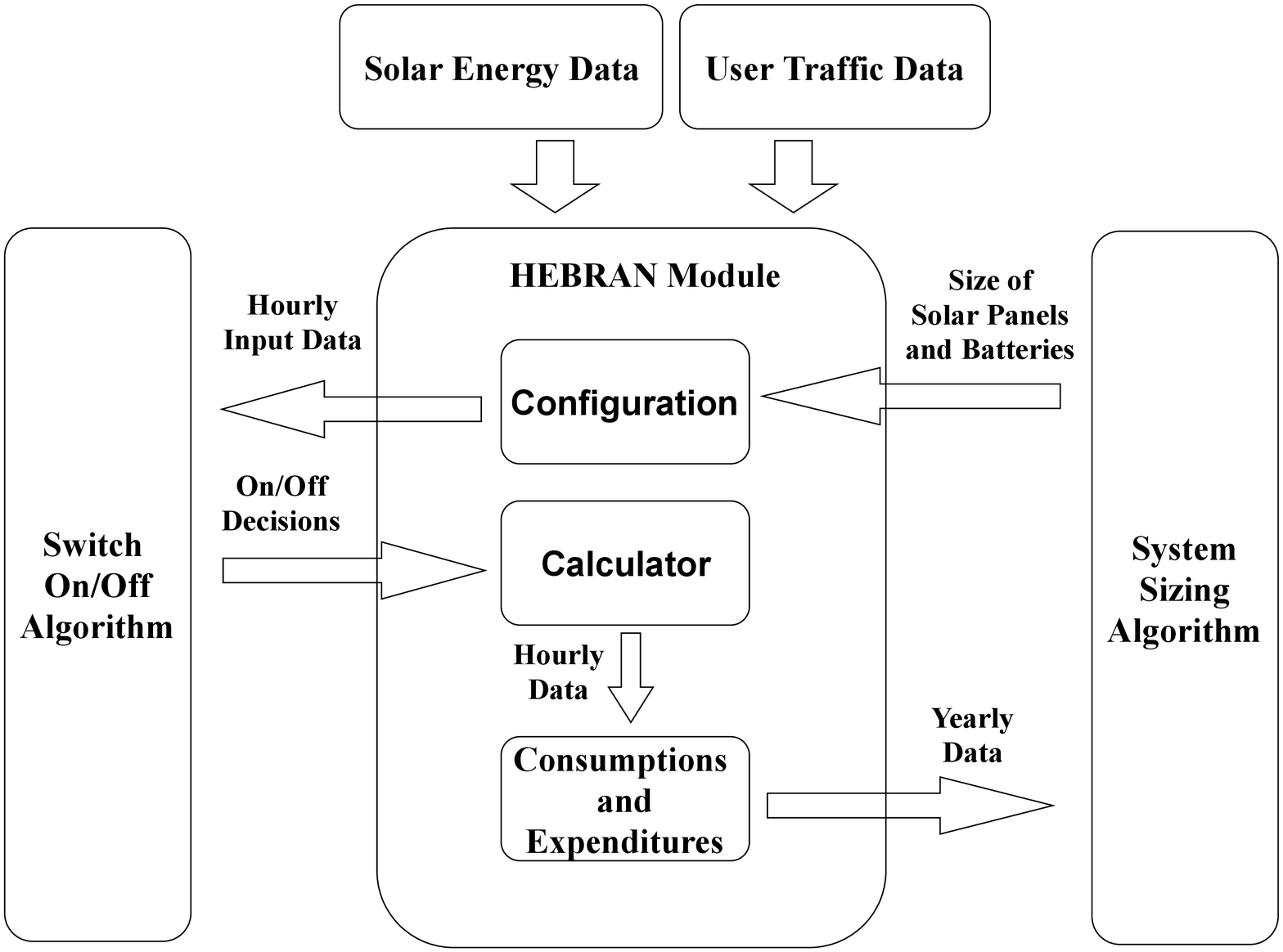}
\caption{\label{fig:framework} HEBRAN Framework }
\end{figure}
\section{Proposed Algorithms for the Optimal Design and Operation of a HEBRAN}
\par In this section, we propose novel algorithms to solve the NP-Complete problem explained in the previous section. First we explain the framework that we use for these algorithms, then give their details. As an alternative for our proposed algorithms, we create a reduced model of the HEBRAN optimization problem to solve it in a Mixed Integer Linear Programming (MILP) Solver, Gurobi \cite{GurobiOptimization2017}. We give the details of this solution in the result section.
\subsection{HEBRAN Framework}
\label{sec:depheur}
\par As we mentioned in the previous section, the solar panels and the batteries are the main investment cost of a mobile operator. The size of these new components should be selected carefully to maximize the economic benefits of the mobile operator. This problem is an offline problem and should be solved before the installation of the panels and the batteries on the base stations. On the other hand, the third component in the objective function (Equation ~\ref{eq:obj1}) is related with the on-grid energy consumption of the base stations. We may reduce this component by using the base station switch on/off algorithms. However these algorithms depend on the data traffic rates and harvesting solar energy by the solar panels, which changes temporally and we can only forecast their values. Therefore this problem is an online decision problem and should be solved for each day with new predicted data. Despite this, the decisions of the offline problem (choosing the size of solar panels and the batteries) directly changes  the renewable energy ratio of the base stations which affects the results of this online decision problem. This relation emphasizes that we have to create a framework, in which the problem is decomposed into the two separate parts. In the first part, we have to focus on an offline algorithm, which aims to choose the size of the solar panels and the batteries for the different traffic rates and harvesting renewable energy rates. In the second part, we have to focus on an online algorithm, which will run during the operation of the system. In addition, we have to use the output data that is created in a part as an input data in the other part. Figure~\ref{fig:framework} illustrates this proposed framework.
\par The harvested renewable energy varies in different months (Figure~\ref{fig:harvested_daily}) and investigating the effect of this variation is one of the purposes of this study. Hence, a HEBRAN operator module is in the core of this framework and it is responsible to operate this network for one year. For this purpose, this module begins with receiving the problem data that are generated with the methods mentioned in Section~\ref{sec:systemDescription}. In the next step, the module runs the system for one year with a switch on/off algorithm. This algorithm makes decisions on which base stations switch off in each time interval. The algorithms used in this module is explained briefly in Section~\ref{sec:heur}. The module receives the decisions from this algorithm and logs each time interval to calculate the energy consumptions and expenditures which are used by the system sizing algorithm (Section~\ref{sec:sizing}).
\begin{algorithm}
\caption{Sizing Solar Panels and Batteries}
\label{alg:sizing}
\begin{algorithmic}[1]
\STATE $\hat{TCO}_{0}=\infty$, $itrt=0$, $step=0$, $fail=0$
\STATE  $\hat{S}_{0}=\hat{1}$,  $\hat{B}_{0}=\hat{1}$
\WHILE{$step<=3$}
\STATE $\hat{TCO}_{itrt+1} \leftarrow$ Run the System for One Year
\IF{ $\hat{TCO}_{itrt+1} > \hat{TCO}_{itrt}$}
\STATE $fail=fail+1$
\IF{$fail>=2$}
\STATE $fail=0$
\STATE $step=step+1$
\ENDIF
\ELSE
\STATE $fail=0$
\ENDIF
\STATE $new\_sizing=False$
\WHILE{$new\_sizing=False$}
\IF{step=0 OR step=2}
\STATE $\hat{S}_{itrt+1}$ $\leftarrow$ Run $ISSP$ ALGORITHM
\IF{$\hat{S}_{itrt+1}=\hat{S}_{itrt}$}
\STATE $step=step+1$
\ELSE
\STATE new\_sizing=True
\ENDIF
\ENDIF
\IF{step=1 OR step=3}
\STATE  $\hat{B}_{itrt+1}$ $\leftarrow$ Run $ISB$ ALGORITHM
\IF{$\hat{B}_{itrt+1}=\hat{B}_{itrt}$}
\STATE $step=step+1$
\ELSE
\STATE new\_sizing=True
\ENDIF
\ENDIF
\ENDWHILE
\STATE $itrt=itrt+1$
\ENDWHILE
\STATE $return \arg\min\limits_{TCO_{itrt}}(\hat{S}_{itrt}, \hat{B}_{itrt}$)
\end{algorithmic}
\end{algorithm}
\subsection{Sizing of Solar Panels and Batteries}
\label{sec:sizing}
\par As we explained in the previous section, deciding the size of the solar panels and the batteries that reduce the TCO of a HEBRAN is one of the main goal of our framework. For this purpose we implemented an offline sizing algorithm (Algorithm~\ref{alg:sizing}), in which we start with the minimum size of solar panels and batteries. Then we create a four-step decision loop that in the first and third step we run the increasing size of solar panels algorithm (ISSP, Algorithm~\ref{alg:issp}), otherwise we run the increasing size of batteries algorithm (ISB,  Algorithm~\ref{alg:isb}). In this loop, we always start with running the system for one year with the online algorithms (Section~\ref{sec:heur}) and calculating the TCO. Then we compare this value with the TCO in the previous iteration. If the TCO increases twice in a row, we switch the type of the increasing algorithm. In addition, we also switch the type of the increasing algorithm if one of the algorithms fails to make a change in the size of panels or batteries which means that it is not feasible to make any more incrementation. Finally in each iteration, we record the sizing configuration ($\hat{S}_{itrt}$,  $\hat{B}_{itrt}$)  and after the breaking of the loop, we return the sizing configuration of the iteration that have the minimum TCO.
\begin{algorithm}
\caption{ISSP Algorithm}
\label{alg:issp}
\begin{algorithmic}[1]
\STATE Given: $itrt$, $\hat{texp}_{itrt}$, $\hat{gexp}_{itrt}$, $\hat{S}_{itrt}$
\FORALL{${i} \in I$}
\STATE $\hat{pot}^{i}=\frac{1}{\hat{S}_{itrt}^{i}}*\hat{texp}_{itrt}^{i}$
\STATE $\hat{pot}^{i}=\min(\hat{pot}^{i}, \hat{gexp}_{itrt}^{i})$
\ENDFOR
\STATE $\hat{pot} \leftarrow sort(\max(\hat{pot}))$
\STATE $max\_count=|I|/2-(itrt*4)$
\STATE $CLIST=\emptyset$
\FOR{$i:=1$ \TO $max\_count$}
\IF{$\hat{pot}^{i}*c^{E}*|$T$|*15>c^{S}$}
\STATE $CLIST=CLIST\bigcup \{i\}$
\ELSE
\STATE break
\ENDIF
\ENDFOR
\STATE $mindist=600$
\STATE $LIST=\emptyset$
\WHILE{$CLIST\neq\emptyset$}
\STATE $LIST=LIST\bigcup CLIST^{0}$
\FORALL{$i\in CLIST$}
\IF{$dist(LIST^{-1}, CLIST^{i})<mindist$}
\STATE $CLIST=CLIST \setminus CLIST_{i}$
\ENDIF
\ENDFOR
\ENDWHILE
\FORALL{$i\in I$}
\IF{$i\in LIST$}
\STATE $\hat{S}_{itrt+1}^{i} =\hat{S}_{itrt}^{i}+1$
\ELSE
\STATE $\hat{S}_{itrt+1}^{i} =\hat{S}_{itrt}^{i}$
\ENDIF
\ENDFOR
\STATE return $\hat{S}_{itrt+1}$
\end{algorithmic}
\end{algorithm}
\begin{algorithm}
\caption{ISB Algorithm}
\label{alg:isb}
\begin{algorithmic}[1]
\STATE Given: $itrt$, $\hat{unstrd}_{itrt}$, $\hat{gexp}_{itrt}$, $\hat{B}_{itrt}$
\FORALL{${i} \in I$}
\STATE  $\hat{pot}^{i}=\min(\hat{unstrd}_{itrt}^{i}, gexp_{itrt}^{i})$
\ENDFOR
\STATE $\hat{pot} \leftarrow sort(\max(\hat{pot}))$
\STATE $max\_count=|I|/2-(itrt*4)$
\STATE $LIST=\emptyset$
\FOR{$i:=1$ \TO $max\_count$}
\IF{$\hat{pot}^{i}*c^{E}*|$T$|*15>c^{B}$}
\STATE $LIST=LIST\bigcup \{i\}$
\ELSE
\STATE break
\ENDIF
\ENDFOR
\FORALL{$i\in I$}
\IF{$i\in LIST$}
\STATE $\hat{B}_{itrt+1}^{i} =\hat{B}_{itrt}^{i}+1$
\ELSE
\STATE $\hat{B}_{itrt+1}^{i} =\hat{B}_{itrt}^{i}$
\ENDIF
\ENDFOR
\STATE return $\hat{B}_{itrt+1}$
\end{algorithmic}
\end{algorithm}
 \par In the ISSP algorithm, first, we sort the base stations according to their energy consumption reducement potential ($\hat{pot}$) if we choose these base stations to increase their panel size for the next iteration ($\hat{S}_{itrt+1}$). The criteria for this potential depends on the total energy consumption ($\hat{texp}$) and grid energy consumption ($\hat{gexp}$) in these base stations in one year period. Then, we limit the number of base stations that can be selected to increase their solar panel size which reduces in each iteration. After this limitation, we check the feasibility of the size incrementation by comparing the reduction in the electricity price (operational expenditure) and the increase in the price of solar panels (capital expenditure). In the last while loop, we also eliminate the base stations which are close to each other and we choose only the base stations which have better potential to decrease the TCO. This elimination prevents unnecessary solar panel incrementation at the base stations in a small area. On the other hand, in the next iterations, in which we run the system with the new size configuration ($\hat{S}_{itrt+1}$), we can choose these eliminated base stations, if these base stations sustain their potential to decrease the TCO. The ISB algorithm is very similar to the ISSP algorithm. However, the potential criteria in this algorithm depends on the unstored\footnote{The unstored energy is the harvested energy by a solar panel but could not be stored in a battery due to fully charged state of this battery.} renewable energy ($\hat{unstrd}_{itrt}$) in the batteries of base stations at the previous iteration.
\subsection{Base Station Switch On/Off Algorithms}
\label{sec:heur}
\par The evidences show that base stations consume very high amount of energy even if they do not provide a service to any location \cite{Auer2011}. This energy consumption arises from non-traffic related operations such as cooling the base station or running the baseband unit subsystem. Therefore, either a load-balancing method or a cell-breathing technique may not reduce the energy consumption of a base station significantly. For that reason, the online algorithms in this paper aim to completely switch off as many as possible base stations in the network.
\par A major problem of a switch on/off algorithm appears on the prediction of the arriving traffic data. The real traffic data from the study of Peng et al. \cite{Peng2014} shows that the traffic patterns of the consecutive days are similar to each other, which means that using the traffic data of the previous day is a good option for the arriving traffic data (Figure~\ref{fig:traffic_dist_day.png}). Despite this, weekend days have different traffic patterns, so it is better to use the previous weekend traffic data for the weekends (Figure~\ref{fig:traffic_dist_week.png}). In our online algorithms we adopt this method to forecast the arriving traffic data for each day. As an alternative forecasting method, a longer historical data may be also collected to predict the arriving traffic data. Peng et al. provide a formula for this purpose in their paper \cite{Peng2014}. Finding the best forecasting method is out of the scope of this paper but it can be easily incorporated into our solution.
\par The online algorithms in this paper are running in the core network. They start from an initial state assuming that each base station in the network switches on to provide a service to any location and the locations are assigned to the base station which provides the best SNR value without violating the maximum utilization value of this base station. This assumption is acceptable for a typical RAN which does not use any energy-efficiency method \cite{oh2011toward}.  We should have noticed that these algorithms run for each time interval separately and for the sake of keeping the notation simple, we suppress the time interval index $t$ in these algorithms.
\begin{algorithm}
\caption{Battery-aware Algorithm}
\label{alg:batteryAware}
\begin{algorithmic}[1]
\STATE Given: $U_{j}$, $S_{ij}$, $z_{ij}$, $R_{i}$, $\forall i\in\mathcal{I}$, $\forall j\in\mathcal{J}$
\STATE Calculate $W_{ij}= U_{j} / S_{ij}$, $\forall i\in\mathcal{I}$, $\forall j\in\mathcal{J}$
\STATE $I^a=\mathcal{I}$  
\STATE $I^s= sort(\arg\min\limits_{i\in\mathcal{I}}(R_{i}))$
\WHILE{$I^s\neq \emptyset$}
\STATE $z_{ij}^P = z_{ij}$, $\forall i\in\mathcal{I}$, $\forall j\in\mathcal{J}$
\STATE $L_{i}= \sum\limits_{j\in\mathcal{J}} W_{ij} * z_{ij}$, $\forall i\in\mathcal{I}$
\STATE $I^a= I^a \setminus  I^s_{1} $
\STATE $J^{I^s_{1}} = \emptyset$
\FORALL{$j\in J$}
\IF {$z_{I^s_{1}j} = 1$}
\STATE $J^{I^s_{1}} = J^{I^s_{1}}  \bigcup \{j\} $
\STATE $z_{I^s_{1}j}=0$ 
\ENDIF
\ENDFOR
\FORALL{$j\in J^{I^s_{1}}$}
\STATE  $I^j = sort(\arg\min\limits_{i\in\mathcal{I}}(W_{ij}))$
\FORALL{$i\in I^j$}
\STATE $L_{i}^N = L_{i} +  W_{ij}$
\IF {$L^N_{i} \leq \rho$}
\STATE $L_{i} = L^N_{i}$
\STATE $z_{ij}=1$
\STATE $break$
\ENDIF
\ENDFOR
\ENDFOR
\IF{$z_{ij}=\emptyset$, $\exists j\in J^{I^s_{1}}$, $\forall i\in\mathcal{I}^a$}
\STATE $z_{ij} = z_{ij}^P$, $\forall i\in\mathcal{I}$, $\forall j\in\mathcal{J}$
\STATE $I^a= I^a \bigcup  I^s_{1} $
\ENDIF
\STATE $I^s= I^s \setminus  I^s_{1} $
\ENDWHILE
\end{algorithmic}
\end{algorithm}
\par In the first algorithm, we aim to switch off the base stations which have the lower renewable energy stored in their battery ($R_{i}$). There are two main energy-efficient benefits in this decision. First, these base stations can increase their stored renewable energy in their battery in the current time interval; thus, the ratio of electrical grid energy usage in these base stations reduces in the following time intervals. Second, the other base stations, which have more renewable energy in their batteries, can spend their renewable energy in the current time interval. Therefore, the probability of the fully charged batteries, which means that we could not store the renewable energy in these batteries, can reduce by activating the base stations which have less reserve capacity in their battery.  (Algorithm~\ref{alg:batteryAware}).
\begin{algorithm}
\caption{Hybrid Algorithm}
\label{alg:hybrid}
\begin{algorithmic}[1]
\STATE Given: $U_{j}$, $S_{ij}$, $z_{ij}$, $R_{i}$, $\forall i\in\mathcal{I}$, $\forall j\in\mathcal{J}$
\STATE Calculate $W_{ij}= U_{j} / S_{ij}$, $\forall i\in\mathcal{I}$, $\forall j\in\mathcal{J}$
\STATE $I^a=\mathcal{I}$  
\STATE $L_{i}= \sum\limits_{j\in\mathcal{J}} W_{ij} * z_{ij}$, $\forall i\in\mathcal{I}$
\STATE $I^s= sort(\arg\min\limits_{i\in\mathcal{I}}(R_{i} + \alpha L_{i}))$
\WHILE{$I^s\neq \emptyset$}
\STATE $z_{ij}^P = z_{ij}$, $\forall i\in\mathcal{I}$, $\forall j\in\mathcal{J}$
\STATE $I^a= I^a \setminus  I^s_{1} $
\STATE $J^{I^s_{1}} = \emptyset$
\FORALL{$j\in J$}
\IF {$z_{I^s_{1}j} = 1$}
\STATE $J^{I^s_{1}} = J^{I^s_{1}}  \bigcup \{j\} $
\STATE $z_{I^s_{1}j}=0$ 
\ENDIF
\ENDFOR
\FORALL{$j\in J^{I^s_{1}}$}
\STATE  $I^j = sort(\arg\min\limits_{i\in\mathcal{I}}(W_{ij}))$
\FORALL{$i\in I^j$}
\STATE $L_{i}^N = L_{i} +  W_{ij}$
\IF {$L^N_{i} \leq \rho$}
\STATE $L_{i} = L^N_{i}$
\STATE $z_{ij}=1$
\STATE $break$
\ENDIF
\ENDFOR
\ENDFOR
\IF{$z_{ij}=\emptyset$, $\exists j\in J^{I^s_{1}}$, $\forall i\in\mathcal{I}^a$}
\STATE $z_{ij} = z_{ij}^P$, $\forall i\in\mathcal{I}$, $\forall j\in\mathcal{J}$
\STATE $I^a= I^a \bigcup  I^s_{1} $
\ENDIF
\STATE $I^s= I^s \setminus  I^s_{1} $
\STATE $L_{i}= \sum\limits_{j\in\mathcal{J}} W_{ij} * z_{ij}$, $\forall i\in\mathcal{I}$
\ENDWHILE
\end{algorithmic}
\end{algorithm}
\par The algorithm starts by calculating the additional system load on a base station if a location is assigned to it ($W_{ij}$). Then, it initialize the active base station set ($I^a$), which includes all the base stations of the network. Next, the algorithm orders the base stations (sorted set $I^S$) by considering the renewable energy stored in their batteries. The order is from the lowest one to the highest one. Finally, the algorithm starts to operate a three-step iteration on this sorted set. 
\par In the first step of the iteration, the algorithm takes a snapshot of the current $z_{ij}^P$ assignment, calculate the current system load ($L_{i}$) in each base station, switches off the first base station in the sorted set (${I^s_{1}}$) and creates an orphaned location set $J^{I^s_{1}}$ which includes the locations that are previously assigned to this switched off base station. In the second step, the algorithm tries to assign these locations to the other base stations in the network without violating the capacity constraints (Equation~\ref{eq:capacityLimit}). This assignment effort is iterated on each active base station ($I^a$) until a base station is found to assign this location. The order starts from the base station which gets the lowest traffic load from this location. In the third step, the algorithm checks the result of the second step and if any uncovered location cannot be assigned to one of the active base stations, it returns the configuration of the network in the beginning of this iteration. This algorithm stops when the sorted set $I^S$ becomes empty. At the end, the algorithm provides a set of base stations which have the highest amount of renewable energy in their battery. In the following time interval, these base stations stay in the active mode and can consume the renewable energy from their batteries which have higher amount of energy than the batteries at switched off base stations.
\begin{figure}
\centering
\includegraphics[width=0.48\textwidth]{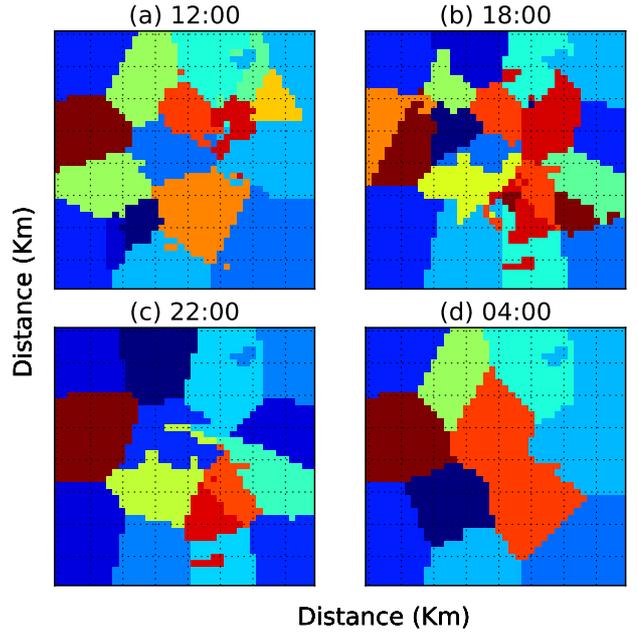}
\caption{\label{fig:assignments} Assignment Decisions of Hybrid Algorithm for Different Time Intervals. }
\end{figure}  
\par The second algorithm called hybrid also considers traffic rates in addition to the battery status. To implement this new adaptation, we have to make two changes in Algorithm 4. First, the algorithm orders the base stations according to a new parameter which is a combination of the remaining renewable energy ($R_{i}$) in the batteries and the system load of a base station $L_{i}$. Second, since the system loads of the base stations change after each switch off decision, the algorithm orders the base stations in each iteration (Algorithm~\ref{alg:hybrid}). Figure~\ref{fig:assignments} shows the results of this algorithm for some selected time intervals.
\section{Computational Experiments}
\label{sec:results}
\par
\begin{figure}
\centering
\includegraphics[width=0.48\textwidth]{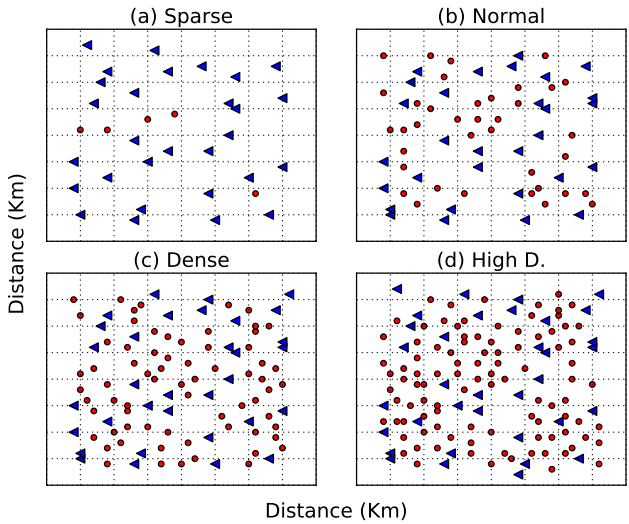}
\caption{\label{fig:bs_locations}The Distribution of Base Stations in Different Traffic Rates. Triangles: Macro Base Stations, Circles: Micro Base Stations}
\end{figure}

 The performance of algorithms are investigated in a $9km^2$ geographical area that is covered by the several macro and micro base stations. The number of the base stations on an area depends on the traffic rate in this area and we repeat the tests for four different traffic rates (Figure~\ref{fig:bs_locations}). In addition, Figure~\ref{fig:harvested_hourly} and Figure~\ref{fig:harvested_daily} show the harvested energy distribution of four different cities we use in the tests to analyze the effect of the solar radiation on our algorithms. In overall, sixteen different combinations were run on this platform (Table~\ref{tab:testChart}). The system used in this analysis is detailed in Section~\ref{sec:systemDescription}. 
\begin{table}
\centering
\caption{\label{tab:testChart} Test Configurations}
\begin{tabular}{|cc|cc| }
\hline
 Traffic Rate &  \# BS &  Solar Radiation &  (kW/year) \\ \hline
 Sparse &  34  &  Stockholm &  986 \\ 
 Normal &  67 &  Istanbul &  1349 \\
 Dense &  102 &  Jakarta &  1359 \\
 High Dense &  134 &  Cairo &  1748  \\
\hline
\end{tabular}
\end{table}
\par Table~\ref{tab:Parameters} shows the test parameters for energy consumptions and the prices of the system. Auer et al. \cite{Auer2011} calculate the value $1350W/h$  and $144.6W/h$  for the energy consumption of a macro base station and a micro base station, respectively. We use these values as our base stations energy consumption ($a^E_{i}$) in full transmission power. In addition, the base stations need to consume energy in the sleep mode for some facilities. This energy consumption does not have any effect on the performance of the algorithms, thus it is omitted in this platform.
\par The system in this paper is expected to operate in the years between 2020 and 2035. Therefore, the price of expenditure costs are calculated for 2020, in which the price of a solar panel is projected to drop to 1\$ per Watt/hour \cite{Munsell2016} and the price of a lithium-ion battery is expected to drop 0.2\$ per Watt/hour \cite{Grothoff2015,Lacey2016}. Lastly, the price of grid electricity is the average price between the 2020 and 2035, in which we also consider the increasing of the price 4\% per annum according to \cite{Energysage2015}.
\begin{table}
\centering
\caption{\label{tab:Parameters} Parameters for Energy Consumptions \& Prices}
\begin{tabular}{|c|c|c| }
\hline
Explanation &  Not. & Value \\ \hline
Energy Cons. of Macro BSs & $a^E_{i}$ &1350 W/h \\
Energy Cons. of Micro BSs & $a^E_{i}$ & 144.6 W/h \\
Unit Cost of a Solar Panel& $c^S$ & 1000\$ \\
Unit Cost of a Battery& $c^B$ & 500\$ \\
Unit Cost of Electricity& $c^E$ & 0.16\$ \\
\hline
\end{tabular}
\end{table}
\par The channel model for macro and micro base stations are chosen as Urban-Macro (UMa) and Urban-Micro (UMi) NLOS cell scenarios from the ITU-R Report \cite{ReportITU-RM.2135-12009}. The value of parameters used to calculate the path loss are listed in Table~\ref{tab:channelParameters}.
\begin{figure}
\centering
\includegraphics[width=0.48\textwidth]{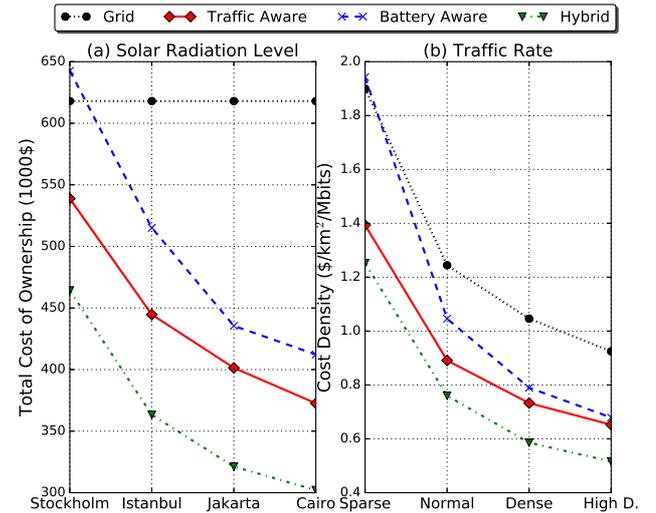}
\caption{\label{fig:bc_cumulative}The Performance of Online Algorithms in Different Solar Radiation Levels and Traffic Rates}
\end{figure}
\begin{table}
\centering
\caption{\label{tab:channelParameters} The Channel Model Parameters}
\begin{tabular}{|c|c|c| }
\hline
Explanation &  Not. & Value \\ \hline
The Carrier Frequency& $f_{c}$ & 1.9 GHz \\
The Channel Bandwidth& $B$ & 20 MHz \\
Street Width& $W$ & 20 m \\
Avg. Build Height& $h$ & 20 m \\
Base Station Height& $h_{bs}$ & 20 m \\
User Equipment Height& $h_{UT}$ & 1.5 m \\
Tx Power of Macro BSs & $P_{i}^T$ & 20 W\\
Tx Power of Micro BSs & $P_{i}^T$ & 6.7 W\\
\hline
\end{tabular}
\end{table}
\begin{figure}
\centering
\includegraphics[width=0.48\textwidth]{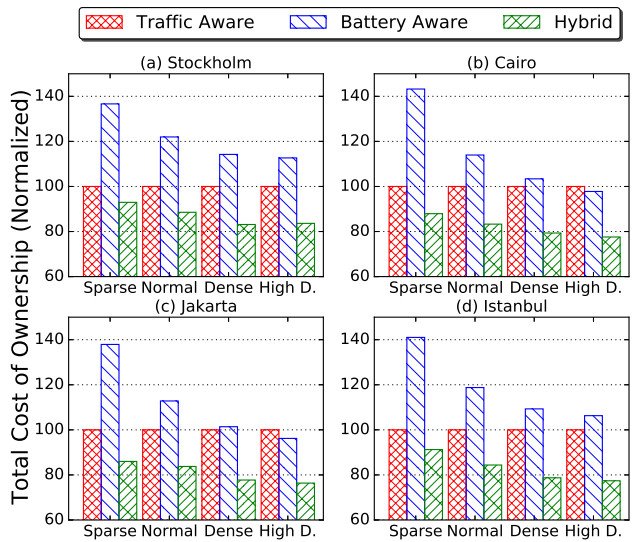}
\caption{\label{fig:bc} Comparison of Online Algorithms in Different Configurations.}
\end{figure}
\par One of the main goals of our design is to find an online algorithm to reduce the electrical grid energy consumption cost. For this purpose we proposed two algorithms in the previous section. We compare these two algorithms with a well-known switch on/off algorithm proposed by Niu et al. \cite{niu2010cell}. The main idea of this algorithm is to switch off as many base stations as possible based on their traffic loads. Figure~\ref{fig:bc_cumulative} shows the performance of the algorithms in different test configurations \footnote{We should noticed that the size of the solar panels and the batteries  in this figure are the best configuration of the sizing heuristic algorithm.}. A number of issues can be identified by this figure. First, the TCO of the system sharply drops with the increasing solar radiation. This is a noticeable result, the more solar radiation provides the more renewable energy generation and yields the lower on-grid energy consumption. Second, the normalized TCO of the system, which is the cost of the system to serve each megabit in one kilometer-square per day, remarkably decrease with the rising traffic rates. This finding provides an important support on using renewable energy sources in urban sectors, in contrast to the common tendency of using renewable energy sources in rural sectors. Lastly, all of the online algorithms perform better than the grid system in any traffic rate and solar radiation combinations except the sparse traffic in Stockholm with battery-aware algorithm. This result shows that investing on a HEBRAN is cost-efficient for different cities around the world which have different traffic rates and solar radiations.
\begin{figure}
\centering
\includegraphics[width=0.48\textwidth]{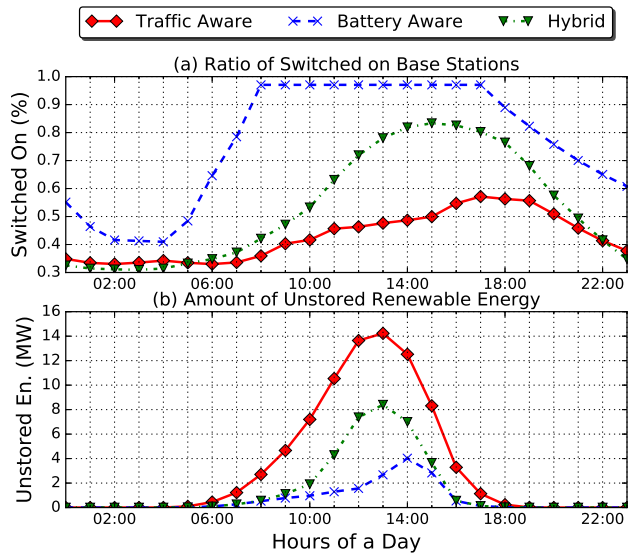}
\caption{\label{fig:active_unstored_hourly} Comparison of Online Algorithms in a Day Period (Istanbul - Sparse Traffic).}
\end{figure}
\begin{figure}
\centering
\includegraphics[width=0.48\textwidth]{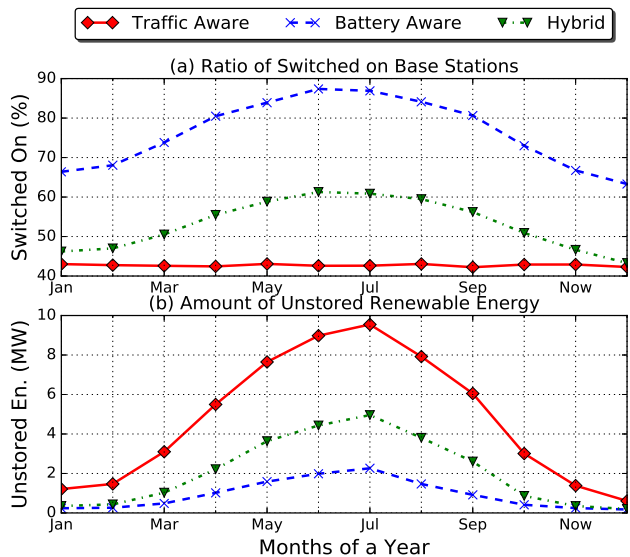}
\caption{\label{fig:active_unstored_monthly} Comparison of Online Algorithms in a Year Period (Istanbul - Sparse Traffic).}
\end{figure}
\par  If we now focus on the comparison of the algorithms, for each solar radiation and traffic rate, the hybrid algorithm is far better than the other two online algorithms. Figure~\ref{fig:bc}, which shows the nominal expenditure of our two algorithms according to the Niu et al. algorithm (traffic-aware algorithm), demonstrates that the hybrid algorithm outperforms the Niu et al. algorithm in any test cases at the range from 10\% to 30\%. While this superior performance is not significantly change with the increasing solar radiation, it gradually increases according to the raising traffic rates. This result yields that our novel algorithm further boosts the idea of using the renewable energy in crowded urban regions.
\par Turning now to the performance of the battery-aware algorithm, Figure~\ref{fig:bc} provides two important findings. First, despite the improving performance in higher traffic rates, it could not reach the performance of the standard and hybrid algorithms. That poor performance can be seen in Figure~\ref{fig:active_unstored_hourly} which shows that this algorithm have higher ratio of switched-on base stations during a day period, which address that the base stations operated with this algorithm could not adapt to the traffic loads efficiently. Second, the battery-aware algorithm provides better performance in the cities that receive more steady solar radiation in different months. The main reason of this result can be demonstrated in Figure~\ref{fig:active_unstored_monthly}. The top subfigure (Fig.~\ref{fig:active_unstored_monthly}a) shows that the variation of the number of switched on base stations between different months is significantly high (more than 20\%) for the battery-aware algorithm, which means that the battery-aware algorithm is more fragile for the change of solar radiation. The subfigure below (Fig.~\ref{fig:active_unstored_monthly}b) shows that in the summer months, the battery-aware algorithm reduces the unstored energy significantly, which is the main benefit of this algorithm,  as we mentioned earlier. In overall, battery-aware algorithm yields some cost benefits due to the efficient battery usage, but it has a lack of adaptation to the changes in solar radiation and traffic rates. To sum up, our hybrid algorithm provides a balance between the adaption to the traffic fluctations and battery utilization, thus it provides better results to reduce the TCO. 
\begin{figure}
\centering
\includegraphics[width=0.48\textwidth]{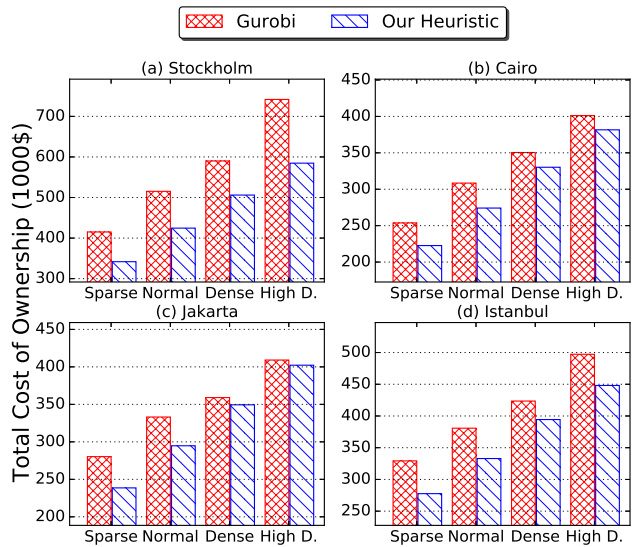}
\caption{\label{fig:gurobi} Comparison of the TCO in our heuristic with a MILP Solver (Gurobi).}
\end{figure}
\begin{figure}
\centering
\includegraphics[width=0.48\textwidth]{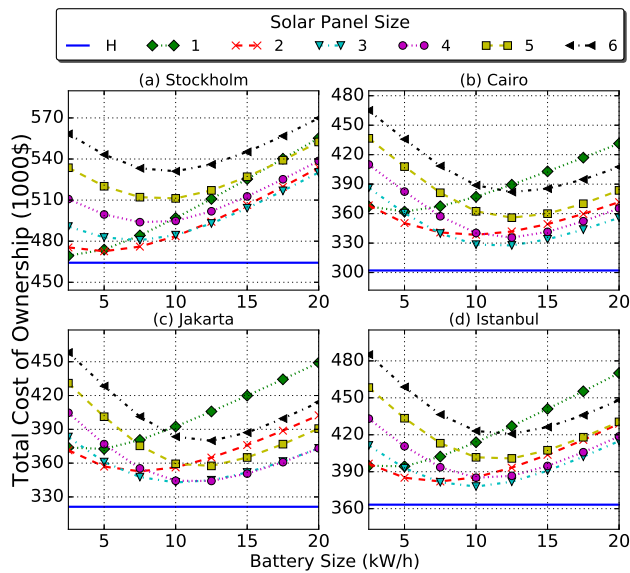}
\caption{\label{fig:combination_per_city} Comparison of the TCO in Different Solar Radiations.}
\end{figure}
\begin{figure}
\centering
\includegraphics[width=0.48\textwidth]{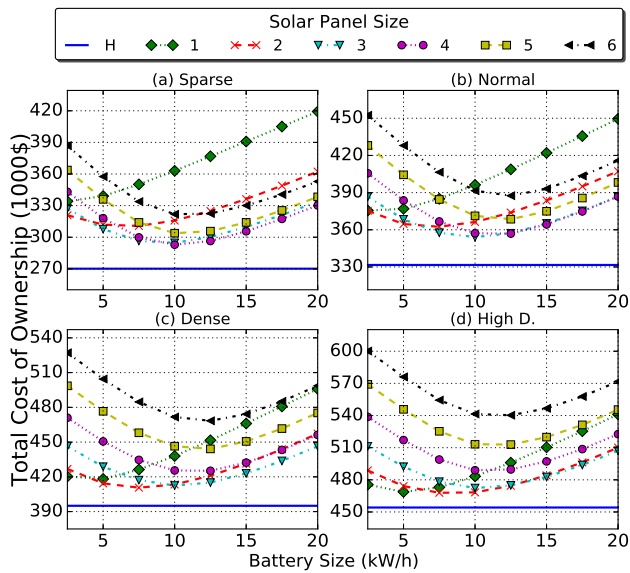}
\caption{\label{fig:combination_per_traffic} Comparison of the TCO in Different Traffic Rates.}
\end{figure}
\par To investigate the performance of our sizing heuristic algorithm, first we compare it with the results of a Mixed Integer Linear Programming (MILP) Solver, Gurobi \cite{GurobiOptimization2017}. As mentioned in Section~\ref{sec:batteryModel} time-coupling property of the original model prevents us to solve it in a MILP Solver. Therefore, we have to use a reduced model in the MILP Solver, in which the base stations should have used their remained renewable energy in a day period. In addition, we have to reduce the number of the time intervals to the 96 (4 days, by using the average solar radiation rates and the traffic rates in a season) to find a solution in a reasonable time in GUROBI. Finally, we run the solver for 14 hours to compare it with our heuristic which find the solution in less than 30 minutes in the same computer\footnote{We should have noticed that all test cases have been executed on a computer with an Intel Xeon E3-1270 Quadcore 3.6GHz processor and 16GB of memory.}. Figure~\ref{fig:gurobi} shows that we can come up with better solutions for all instances we studied\footnote{As it was mentioned before, the hybrid algorithm outperforms the other two algorithm in any test case. Therefore we demonstrates only the sizing results that use the hybrid algorithm as an online algorithm.}.
\par Our sizing heuristic also outperforms the systems with the same size solar panels and batteries in any solar radiations rate (Figure~\ref{fig:combination_per_city}) and the traffic rate (Figure~\ref{fig:combination_per_traffic}). Those results can encourage the MNOs employ our heuristic with the historical solar radiation and traffic data to decide on the size of solar panels and the size of batteries to reduce their TCO. On the other hand, these two figures provide us valuable information about the correlation between the size of solar panels and cities/traffic rates. For example in Stockholm, where the solar radiation rate is very low and its variation is very high (Figure~\ref{fig:harvested_daily}), the smaller solar panels provides more profit to a MNO (Figure~\ref{fig:combination_per_city}). However, we could not find a strong correlation between the increasing solar radiation and the solar panel sizes, if we compare the results of the other cities. Another result is that the smaller solar panels are more cost-efficient in the test cases that have higher traffic rates. The main reason of this outcome is the higher amount of base stations in these sectors, thereby increasing of capital expenditure. This result can be seen more easily in Figure~\ref{fig:combination_per_traffic}, which shows the TCO according to the mean values\footnote{Therefore we can bypass the effect of solar radiation rate on the results.} of four cities. In the subfigures below (higher traffic rates), the offsets between the consecutive solar panel sizes are larger than the top subfigures, which means that the reducing amount of the operational expenditure could not compensate by the increasing amount of the capital expenditure in the higher traffic rates.

\section{Conclusion}
\par Increasing electrical energy costs enforces the mobile network industry to focus on energy efficient solutions. In addition, reducing the carbon emission rates is an emerging issue for RANs. In this paper, we describe a new type of RAN, in which the base stations in this network have a connection to the electrical grid and they have their own solar panel and battery. We formulate an optimization problem which aims to reduce the TCO in this network and propose a framework for the solution. In our solution, we describe several algorithms that target to find both the ideal solar panel and battery sizes in the base stations and the ideal schedules for these base stations to reduce the electrical grid cost of the MNO.
\par The results show that our sizing algorithm is more cost-efficient than the MILP solver in all cases we studied. This algorithm also outperforms the systems in which the size of solar panels and batteries are identical. Our hybrid switch on/off algorithm manages the renewable energy in the batteries of the base stations in an efficient way and get benefits from the data traffic variation in a RAN to reduce the on-grid electricity. Interestingly, the results show that a HEBRAN provides better results with the increasing traffic rates. In conclusion, our framework and proposed algorithms reduce the TCO and provide an environmental network; thus, they support the usage of the renewable energy sources by a MNO especially in an urban sector of a city.
\par As a future work, we are planning to adopt more detailed cost and renewable energy system models to our framework. In addition, we will remodel our problem and find the solutions to reduce the carbon emission rates in a HEBRAN. Lastly, our framework and algorithms may be migrated to the next generation RAN architectures such as a Hybrid-CRAN. Splitting the digital units between the different levels of a Hybrid-CRAN is investigated broadly for bandwidth-efficiency and delay-efficiency in recent years. However, using the renewable energy systems cost-efficiently in this architecture is an open issue and finding a solution can reduce the TCO of a MNO.  
\def\bibsection{\section*{References}}
\bibliography{journal1}

\end{document}